\documentclass[a4paper,11pt]{article}
\usepackage{jinstpub} 
\usepackage{lineno}

\title{\boldmath Anti-scatter grid prototype manufactured via laser powder bed fusion of pure tungsten}

\author[a]{I. Diachkov,\note{Corresponding author.}}
\author[a]{S. Chernyshikhin,}
\author[a]{S. Dadabaev,}
\author[a]{S. Kholodenko,}
\author[b]{A. Kulik,}
\author[c]{V. Obraztsov,}
\author[a]{E. Shmanin,}
\author[d]{D. Strekalina}

\affiliation[a]{National University of Science and Technology <<MISIS>>,\\ 119049, Leninskiy Prospekt 4, Moscow, Russia}
\affiliation[b]{Institute for Nuclear Research of Russian Academy of Sciences, 117312 Moscow, Russia}
\affiliation[c]{NRC <<Kurchatov Institute>> - IHEP\\ 142281, 1 Nauki sq., Protvino, Russia}
\affiliation[d]{Moscow Polytechnic University\\ 38, Bolshaya Semyonovskaya, Moscow, 107023, Russia}
\emailAdd{diachkov.is@misis.ru}

\abstract{
This paper describes the production and tests performed with the tungsten collimator prototype produced using laser powder bed fusion technology in a converging geometry. This prototype has been tested for being the anti-scatter grid for X-ray imaging to reduce background from the scattered photons. The purpose is reducing patient radiation exposure in medical applications. Obtained results are positive, grid provides definitive enhancement in marker detection which proves to be especially useful with artificially created obstacles to imitate harsher working environments. 

}


\begin{document}
\maketitle
\flushbottom

\section{Introduction}
\label{sec:introduction}

Anti-scatter grids are widely used in medicine for X-ray imaging~\cite{Aichinger2004}. The idea was originally proposed by Gustav Peter Bucky in 1913, suggesting that an absorbing grid placed between the patient and the detector would reduce the number of scattered X-rays, improving the quality of the image~\cite{10.1159/000405783}. Lead is commonly used as an absorbing material, but it has several drawbacks~\cite{LEHMANN2002202}. 
With the development of tungsten laser powder bed fusion (LPBF) technology, tungsten is becoming the preferred choice. 
Tungsten's suitability for anti-scatter grids stems primarily from its remarkable X-ray attenuation capability~\cite{Behling2018DiagnosticXS}. 
With a high atomic number of 74, tungsten effectively interacts with X-rays, absorbing and blocking scattered radiation. 
Unlike lead, tungsten exhibits remarkable mechanical strength and resilience to wear and tear.
This durability ensures that anti-scatter grids maintain their effectiveness over time, providing consistent image quality throughout their lifespan~\cite{REN2018170}. 

\begin{figure}[h]
    \centering
    \includegraphics[width=0.5\linewidth]{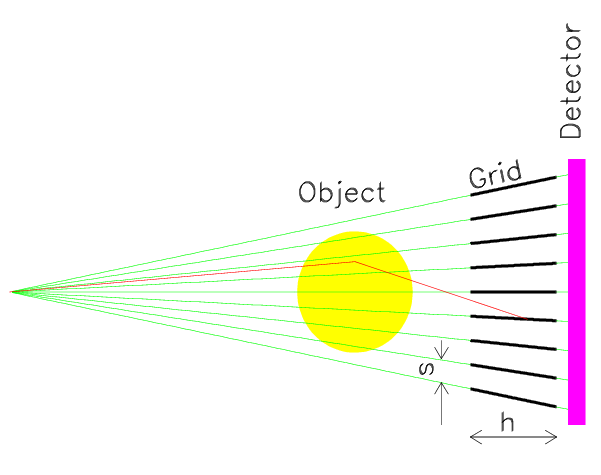}
    \caption{Illustration of the test setup and anti-scatter grid geometrical properties.}
    \label{fig:sketch}
\end{figure}

Collimators with parallel walls and converging geometry can be used. In the latter case grid walls are designed to be aligned with the X-rays from the point-like source (Figure \ref{fig:sketch}). Such design has an advantage of reduced absorption of primary X-rays due to smaller cross-section with collimator walls. However, production of such collimator is non-trivial. The specimens currently available on the market have very thick walls.

In this paper, a tungsten prototype produced by laser powder bed fusion of pure tungsten is described. The LPBF technology was developed in the National University of Science and Technology <<MISIS>> for the production of detectors for high-energy physics. Using that technology, collimators with converging geometry can be produced. The durability of the tungsten also allows the walls to be thinner while the high density leads to significant absorption coefficient of refracted X-rays. These factors combined reduce the absorption of primary X-rays. As a result, time needed to collect enough data and therefore the radiation exposure of the patient are reduced~\cite{Desai01112020}. The usage of converging geometry requires precise positioning and alignment of the grid with respect to the X-ray source. It is not possible to move the grid during data collection in order to smear the collimator shadows across the full image, but rotation is possible.

\section{Manufacturing}

\begin{figure}[htbp]
\centering
\includegraphics[width=.9\textwidth]{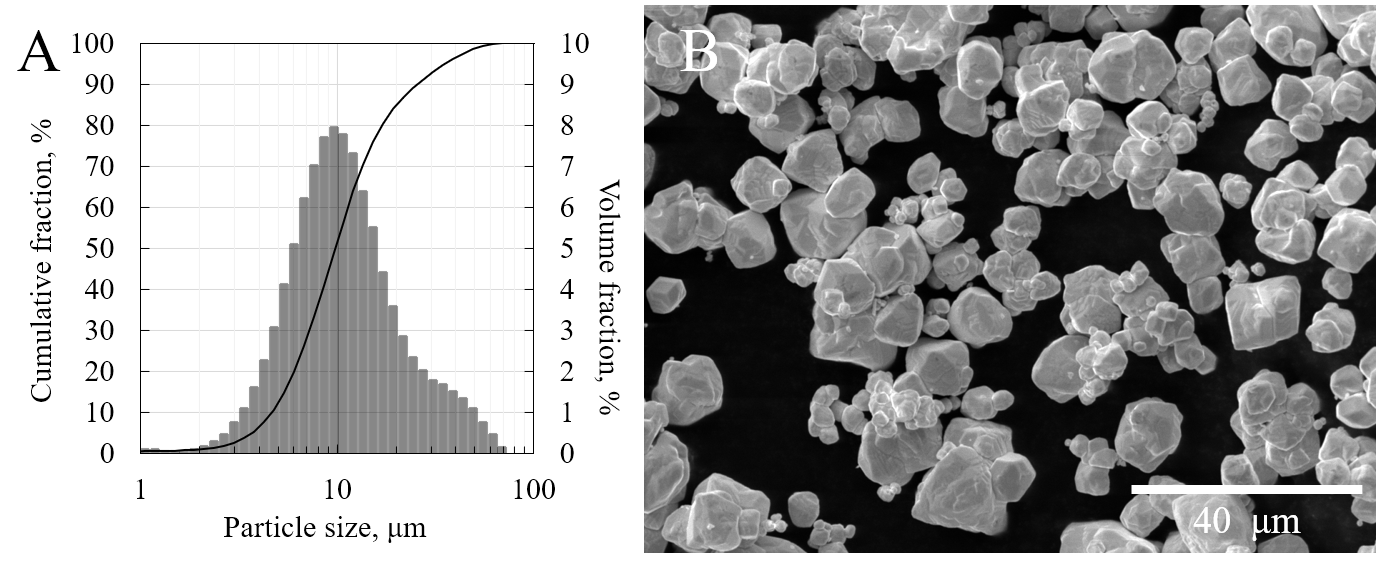}
\caption{Characteristics of tungsten powder utilized for LPBF process: particle size distribution (left), SEM image of particles (right).\label{fig:LPBF}}
\end{figure}

Tungsten W1 powder produced by EOS GmbH (Germany) was utilized as a raw material for anti-scatter grid manufacturing. The granulometric composition of powder was carried out on ANALYSETTE 22 Nanotec (Fritsch, Germany) laser particles sizer. The morphology of powder particles was investigated with a scanning electron microscope (SEM) Vega 3 (Tescan, Czech Republic). 
SEM image of tungsten powder is depicted in Figure \ref{fig:LPBF} (left). The particles exhibit mostly polyhedron morphology with an insignificant number of satellites. The particle size had a normal unimodal distribution as shown in Figure \ref{fig:LPBF} (right). The equivalent diameter percentiles were $d_{10} = 4.6\, \mu m$, $d_{50} = 9.7\, \mu m$, and $d_{90} = 25.6\, \mu m$. The flow rate of 108 seconds per 600 grams was obtained on the Hall funnel with a hole diameter of 4 mm. Thus, even though the powder has a non-spherical shape and a relatively small median particle size, it satisfies the LPBF process requirements for an adequate flow rate to create a dense and uniform layer.

The LPBF process was performed on Addsol D50 machine (Additive Solutions, Russia). The LPBF setup was designed for experimental purposes, featuring a relatively small cylindrical building volume with a height of 150~mm and a circular base plate of 50 mm in diameter. A steel base plate was utilized as a substrate for the initial layers of tungsten grid. The machine is equipped with a 400 W fiber laser (IPG Photonics, Russia). Laser with Gaussian power density distribution and 80 $\mu$m laser spot operates at a wavelength of 1070 nm. The process was conducted in an argon atmosphere with an oxygen content below 100 ppm. The executable files were created using Glicer software (ATSS, Russia). The thin-walled grid was created using a single-track-per-layer approach, indicating the highest possible resolution. The main process conditions were as follows: laser power of 350 W, a scanning speed of 1200 mm/s, and a layer thickness of 20 $\mu$m. The scanning speed significantly affects the size of the melt pool, thus at preliminary experiments, the scanning speed was increased to the maximum value (1200 mm/s) at which the wall integrity (continuity) was maintained. In previous work it was demonstrated that the layer thickness of 20 $\mu$m result in higher mechanical properties \cite{CHERNYSHIKHIN2024106699}. After the LPBF process, the samples were separated from the substrate using a GX-320L electrical discharge machine (CHMER EDM, China).

The height and step of the periodic structure are 50~mm and 5~mm respectively, while the designed wall thickness is ~100~$\mu$m. The grid prototype presented in the study consists of 6x6 cells.
The limitation of the LPBF machine available in the laboratory determines the prototype's outer dimensions.


\section{Experimental setup}\label{sec:experimental_setup}

\begin{figure}[h]
    \begin{minipage}{0.38\linewidth}
    \includegraphics[width=\linewidth]{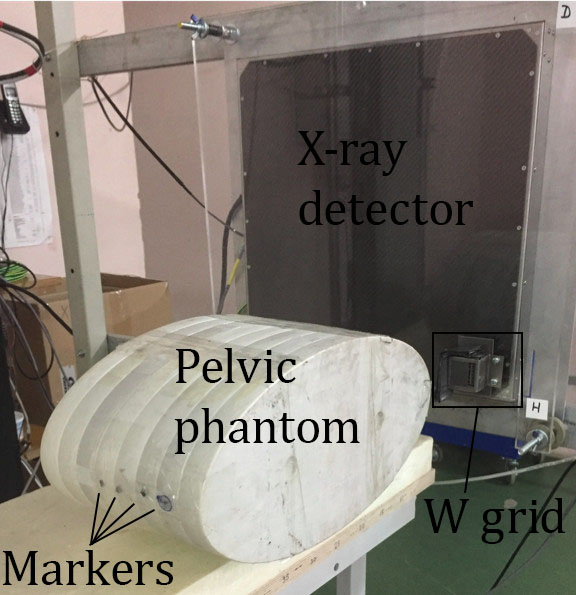}
    \end{minipage}
    \begin{minipage}{0.57\linewidth}
    \includegraphics[width=\linewidth]{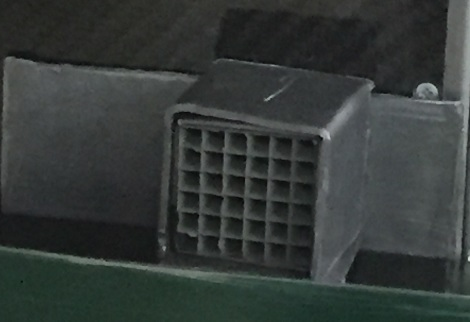}
    \end{minipage}
    \caption{Experimental setup consisting of pelvic phantom with set of grains of a different density and X-ray detector (left) and a converging collimator prototype with surrounding shielding (right).}
    \label{fig:Phantom body}
\end{figure}

The tests of the produced tungsten grid were conducted using a special setup schematically shown in Figure~\ref{fig:sketch}. It consists of the X-ray source, a 40 cm thick human pelvic phantom equipped with an additional set of grains of different density, the prototype of the converging collimator and a detector-sensitive plane~(Figure~\ref{fig:Phantom body}). The X-ray source is located at a 2~m distance from the detector-sensitive plane. The converging collimating grid is located behind the test object in front of the detector plane.

\begin{figure}[h]
    \begin{minipage}{0.48\linewidth}
    \includegraphics[width=\linewidth]{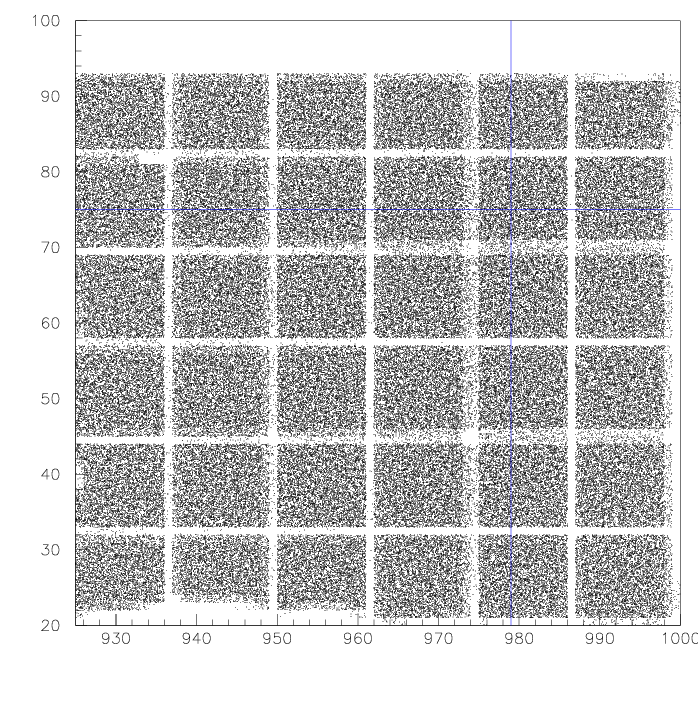}
    \end{minipage}
    \begin{minipage}{0.48\linewidth}
    \includegraphics[width=\linewidth]{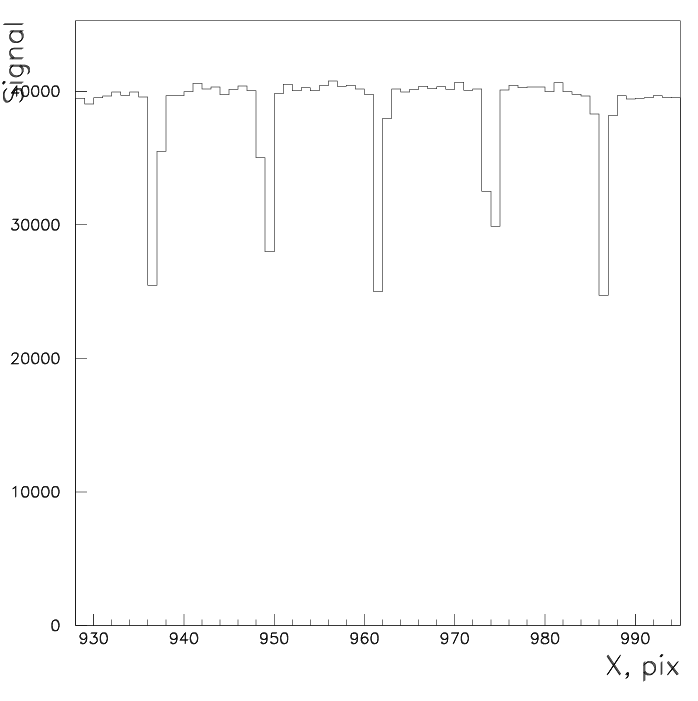}
    \end{minipage}
    \caption{X-ray image of the collimating prototype installed in the bottom right corner(left). The projection on the $X$ coordinate (right)}
    \label{fig:adjustment}
\end{figure}

The small size of the prototype requires precise alignment. In Figure~\ref{fig:adjustment} (left), the image of the grid is presented as a 2D histogram of the signal amplitudes vs $X$ and $Y$ coordinates. Coordinates are measured in the pixel ID. The right figure shows the 1-dimensional projection dependence on the $X$ coordinate. The periodic structure with the reduced signal corresponds to the shadows of the collimator. The signal does not drop down to zero, which means that the pixels (~ 420 $\mu$m in size ) are only partially shaded. Often, 2 pixels are affected. This is inevitable if the collimator period is not precisely aligned with the detectors' pixel positions. The signal integral value is used as the adjustment criteria, i.e. the sum of the shadows in all pixels in a given dip. This value is normalized to the signal integral in the clear part of the image outside the dip (40,000 channels). 
\par
\begin{figure}[h]
    \begin{minipage}{0.48\linewidth}
    \includegraphics[width=\linewidth]{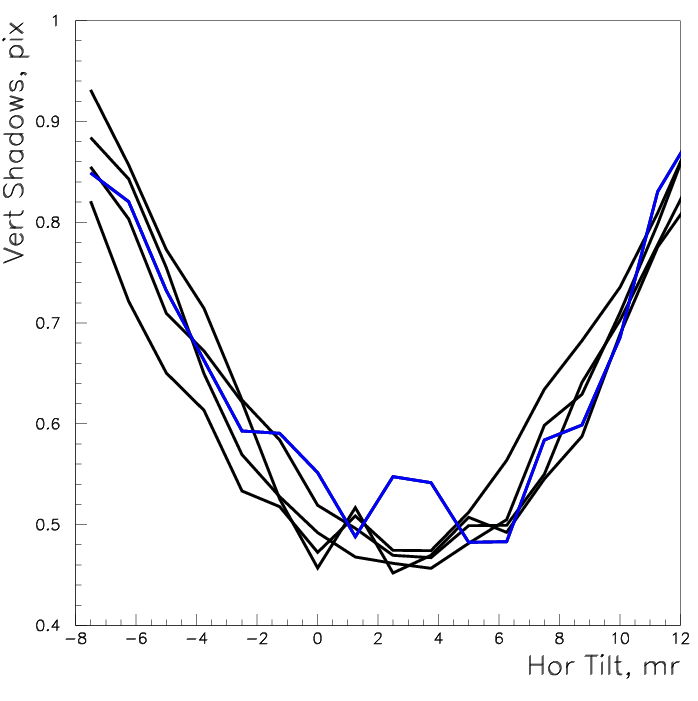}
    \end{minipage}
    \begin{minipage}{0.48\linewidth}
    \includegraphics[width=\linewidth]{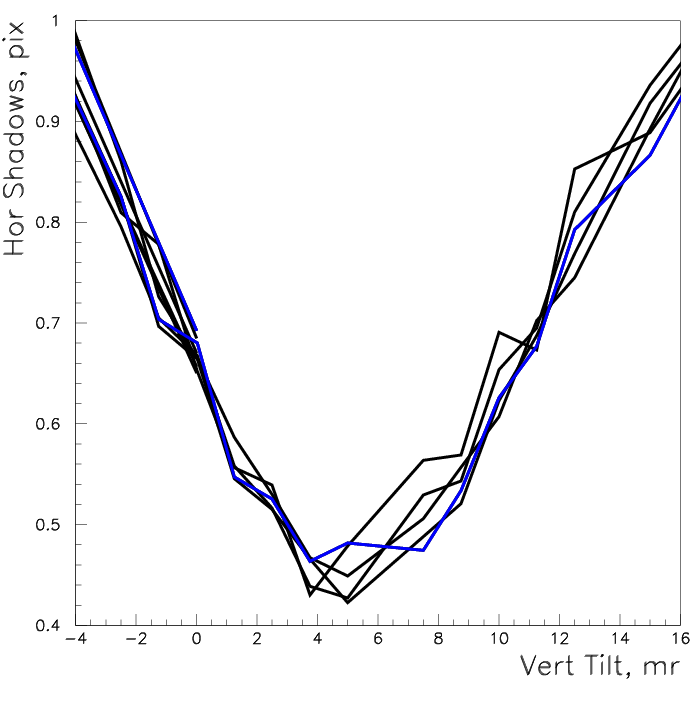}
    \end{minipage}
    \caption{ADC signal integral as a function of prototype tilting angle. }
    \label{fig:angle_adjustment}
\end{figure}

 Five different integral values are monitored by rotating the prototype. Figure~\ref{fig:angle_adjustment} shows the defences of all five integrals as a function of the prototype horizontal (left) and vertical~(right) inclination angles. The minimum value defines the working point. The minimum value is less than half a pixel size, which puts the estimate of the collimator wall effective thickness below 200~$\mu$m.

\section{Data analysis}
\label{sec:analysis}

To estimate the effect of the anti-scatter grid on the quality of the image, several series of experiments were conducted.
The test objective was to compare the clarity of the thick object image with and without the anti-scatter grid. For that purpose the images of the pelvic phantom with several 2~mm lead markers placed on its surface were taken. The measure of success was the ability to detect and measure these markers in different configurations. To take a control image without grid the phantom was lifted up instead of physically removing the collimator. Due to the small prototype grid size, even after small vertical displacement of the object, the marker projection moves to the detector area not affected by the grid. This allows fast switching between “grid” and “no-grid” modes without doing careful collimator alignment every time. Therefore, images with and without grid have different marker pixel coordinates.

\begin{figure}[htbp]
\centering
\includegraphics[width=.7\textwidth]{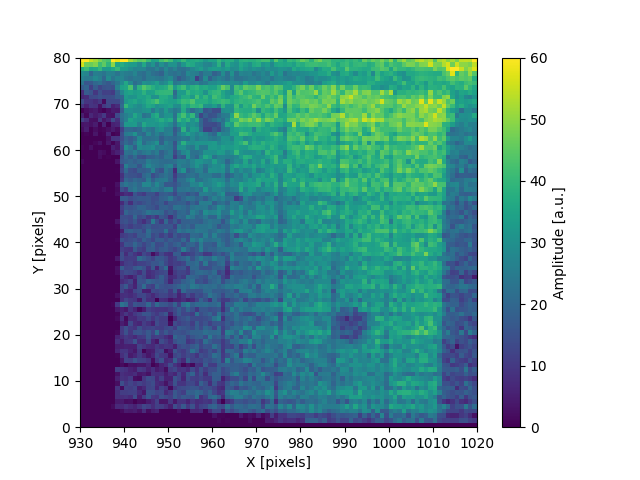}

\caption{Image of the grid and the markers. \label{fig:grid}}
\end{figure}

The image taken with the grid and markers is shown in Figure \ref{fig:grid}. The shadow of the collimator is clearly visible with a period of 12.5 bins between walls. However, periodic structure can be removed using Fourier transform. This method is widely used in image processing to highlight important features or improve image quality. Images represented in all subsequent figures are zoomed in to the relevant region containing the markers and processed identically using Fast Fourier Transform (FFT) algorithm \cite{fft_paper}.

\begin{figure}[htbp]
\centering
\includegraphics[width=.47\textwidth]{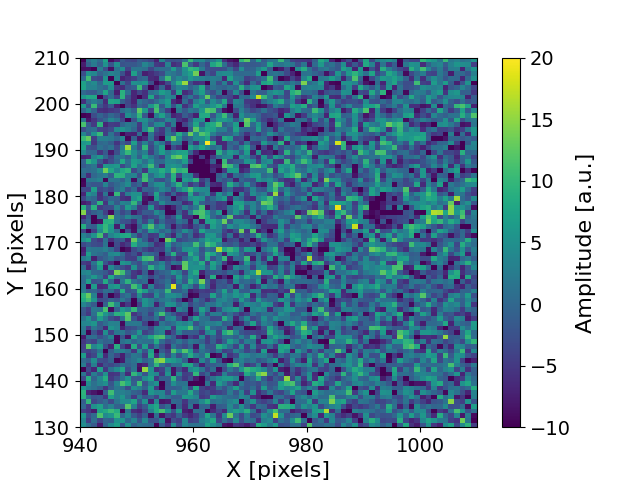}
\qquad
\includegraphics[width=.47\textwidth]{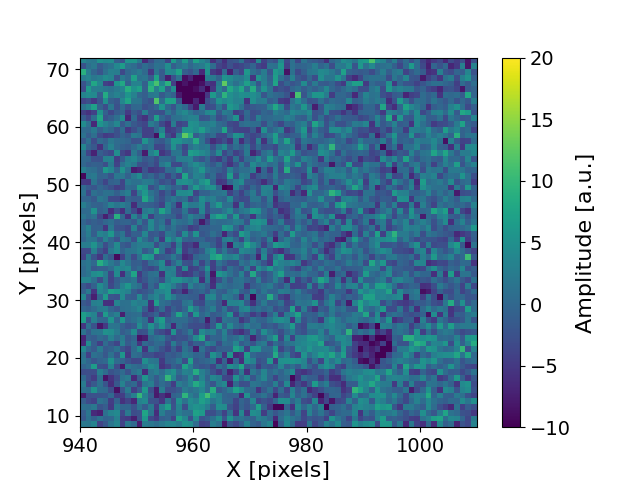}
\caption{Comparison of the image without (left) and with (right) grid assistance. The images are processed using FFT algorithm. \label{fig:no_plex}}
\end{figure}

The results of the experiment with and without the grid are shown in Figure \ref{fig:no_plex}. The shadows left by markers are clearly visible, and their size could be measured. The shadow of the grid is successfully subtracted from the image using FFT technique. Both the collimator prototype and FFT algorithm do not lead to any apparent distortion or degradation of the image. Moreover, less noise is observed in the image taken with grid assistance. The change in the angle between two markers is due to the movement of the pelvic phantom.

\begin{figure}[htbp]
\centering
\includegraphics[width=.47\textwidth]{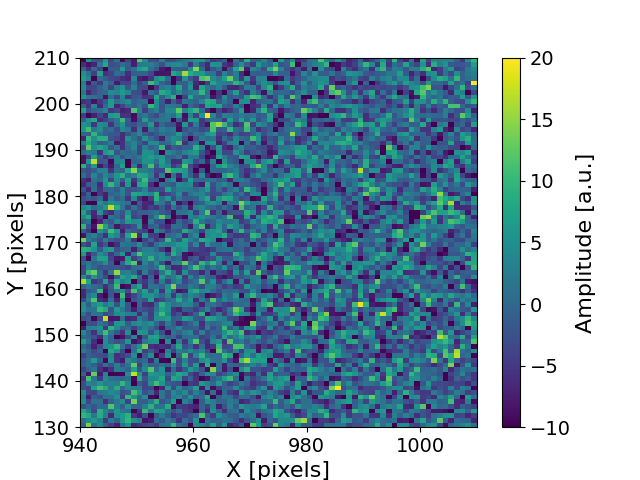}
\qquad
\includegraphics[width=.47\textwidth]{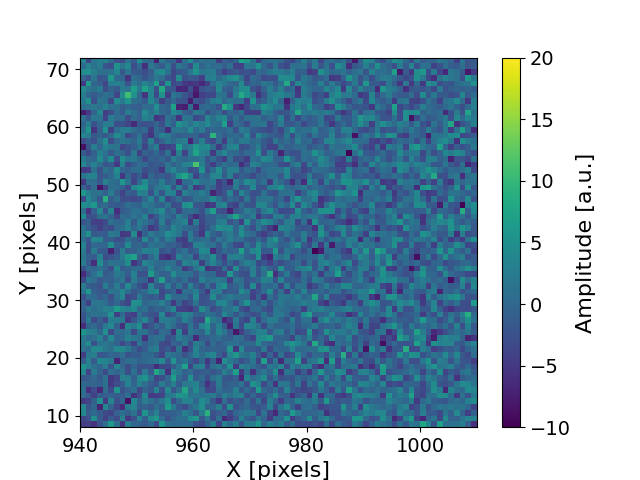}
\caption{Comparison of the image without (left) and with (right) grid assistance. 50 mm plexiglass layer is placed in front of the detector. The images are processed using FFT algorithm. \label{fig:plex}}
\end{figure}

Clearly enough, the anti-scatter grid is expected to offer significant image improvement only for a thick object since thin objects provide less material for X-ray scattering and are imaged well even without grid. Therefore 50~mm thick plexiglass plate was placed in front of the object to increase it's thickness in order to demonstrate the capability of the technique under harsh conditions. Figure \ref{fig:plex} demonstrates the effect this shielding has on the image quality.

Here the benefit of using the collimator becomes more obvious. The image on the left consists only of background noise with the marker shadows being almost indistinguishable. The image on the right includes two marker shadows clearly visible to the naked eye.

\section{Conclusions}
\label{sec:conclusions}

Anti-scatter grid prototype for X-ray imaging has been manufactured with the LPBF technology and tested using the X-ray source, the  human pelvic phantom with additional markers and X-ray detector. Tungsten LPBF technology allows to produce collimator with converging geometry, which results in smaller cross-section of grid walls with direct X-rays from point-like source. That leads to reduced patient radiation exposure compared to traditional collimators used in medicine. 

The results of the performed experiments demonstrate the effectiveness of anti-scatter grid in background reduction and visual marker recognition. While using the collimator makes the image clearer in idealized conditions, it proves to be crucial in distinguishing markers when some kind of additional obstacle is present. The shadows of thin tungsten walls are clearly visible, but do not obstruct markers of this size. These shadows can be subtracted from the image using FFT.

The dimensions of the current prototype are defined by the limitations coming from the LPBF machine used. While scaling up the grid in order to be able to scan larger portions of patient's body is necessary, it is not expected to change performance significantly.
Further experiments are required in order to develop anti-scatter grids with converging geometry and use them for medical applications.


\acknowledgments

We express our gratitude to ZAO Protom company for providing their facility for testing.


\bibliographystyle{JHEP}
\bibliography{biblio.bib}


\end{document}